\documentclass[conference]{IEEEtran}
\usepackage{microtype}
\usepackage{type1cm}
\usepackage{url}
\usepackage{amsmath}
\usepackage{amsthm}
\usepackage{amssymb}
\usepackage{bm}
\usepackage{psfrag}
\usepackage{graphicx}
\usepackage{xspace}
\usepackage{xcolor}
\usepackage{float}
\usepackage[caption=false,font=footnotesize]{subfig}
\usepackage{cite}
\usepackage{skull}
\IEEEoverridecommandlockouts

\DeclareMathOperator*{\argmin}{arg\,min}

\newcommand{\thrp}{\mathsf{S}}
\newcommand{\pwr}{\mathcal{E}}
\newcommand{\btcs}{\mathsf{B}}

\begin{document}
\title{Throughput, Bit-Cost, Network State Information:\\Tradeoffs in Cooperative CSMA Protocols}

\author{
\IEEEauthorblockN{Georg Bocherer and Rudolf Mathar
}
\IEEEauthorblockA{\IEEEauthorrefmark{1}Institute for Theoretical Information
Technology\\
RWTH Aachen University,
52056 Aachen, Germany\\ Email: \{boecherer,mathar\}@ti.rwth-aachen.de}
\thanks{This work has been supported by the UMIC Research Centre, RWTH Aachen University.}
}
\maketitle

\begin{abstract}
In wireless local area networks, spatially varying channel conditions result in a severe performance discrepancy between different nodes in the uplink, depending on their position. Both throughput and energy expense are affected. Cooperative protocols were proposed to mitigate these discrepancies. However, additional network state information (NSI) from other nodes is needed to enable cooperation. The aim of this work is to assess how NSI and the degree of cooperation affect throughput and energy expenses.
To this end, a CSMA protocol called fairMAC is defined, which allows to adjust the amount of NSI at the nodes and the degree of cooperation among the nodes in a distributed manner. By analyzing the data obtained by Monte Carlo simulations with varying protocol parameters for fairMAC, two fundamental tradeoffs are identified: First, more cooperation leads to higher throughput, but also increases energy expenses. Second, using more than one helper increases throughput and decreases energy expenses, however, more NSI has to be acquired by the nodes in the network. The obtained insights are used to increase the lifetime of a network. While full cooperation shortens the lifetime compared to no cooperation at all, lifetime can be increased by over $25\%$ with partial cooperation.
\end{abstract}

\section{Introduction}

Low cost and ease of deployment of wireless local area networks (WLAN) are partially
due to the use of a simple \textit{distributed} medium access control (MAC) protocol. The distributed coordination function (DCF) in IEEE 802.11 guarantees asymptotically the same fraction of channel occupation to each node in the network \cite{Bianchi2000}.
Additionally, multirate capabilities of IEEE 802.11 have enabled WLAN hotspots
to serve nodes with different channel conditions simultaneously. In the uplink,
different channel conditions however result in a strong discrepancy of the
experienced performance among the nodes depending on their
location \cite{Heusse2003}.

Cooperation in wireless networks is a promising approach to mitigate this performance discrepancy between nodes in wireless networks. In \cite{Sendonaris2003}, the authors illustrated that cooperation between two nodes can be beneficial for both nodes, under the assumption of perfect time division multiple access. Distributed protocols were proposed to coordinate cooperation at the MAC layer, for instance
\textit{r}DCF~\cite{Zhu2006a} and CoopMAC~\cite{Liu2007}. Both protocols enable
two-hop transmission as an alternative to direct transmission for WLAN. These
protocols also coordinate cooperation on the physical layer~\cite{Liu2008}.
The benefits of cooperation for the whole network have been discussed
in~\cite{Korakis2007a,Narayanan2007}. In~\cite{Zhu2006a,Bletsas2006,Liu2007}, the authors proposed to optimize the throughput through an appropriate relay selection for each transmission separately.

Although cooperation has the potential to increase throughput, a cooperative network is more demanding  than a non-cooperative network: first, the coordination of channel access is more complex, second, cooperating nodes need more Network State Information (NSI), i.e., information about the other nodes, third, nodes willing to help other nodes in their transmissions have higher energy expenses. In this work, we use a simple model to parameterize these demands. To take the coordination problem into account, we consider a network where channel access is coordinated by carrier sense multiple access (CSMA) and define fairMAC, a parameterizable cooperative CSMA protocol of which we presented a preliminary version in \cite{Bocherer2008}. In fairMAC, the amount of NSI and the degree of cooperation is adjustable. We apply fairMAC with varying parameters to a network of $32$ randomly distributed nodes that all transmit data to a common sink and measure the resulting throughput and the resulting energy expenses. The analysis of the data reveals two fundamental tradeoffs in cooperative networks: first, more cooperation means higher throughput but also higher energy expenses. Second, more NSI leads to higher throughput and lower energy expenses but requires that the nodes perform network discovery.
The insights obtained from the analysis are used to maximize the lifetime of a network. Compared to no cooperation at all, full cooperation decreases lifetime while lifetime can be increased by over $25\%$ by cooperating only partially.

The remainder of this paper is organized as follows. In Section~\ref{sec:model}, we define our system model. We then motivate our work in Section~\ref{sec:keyidea} by considering the throughput/bit-cost tradeoff in a simple example. In Section~\ref{sec:protocol}, we define the parameterizable protocol fairMAC. We finally discuss the tradeoffs that can be adjusted by fairMAC in Section~\ref{sec:simulation}.

\section{System Model}
\label{sec:model}
We consider a network of $N$ nodes that seek to transmit their data to a common
access point (AP). For each pair of nodes $k,l$ of the network, we associate with the transmission from $k$ to $l$ the achievable rate $R_{kl}$. We denote by $R_k$ the achievable rate for direct transmission from node $k$ to the AP. We aim to guarantee the same
throughput to all nodes in the network, irrespective of their achievable transmission rates, therefore, we constantly set the amount of data per packet to $1$~nat. The packet length for a
transmission from node $k$ to the AP is then given by $1/R_{k}$. 
For the cooperative protocols CoopMAC and fairMAC (to be
presented in Section~\ref{sec:protocol}), some nodes have the possibility to transmit their packets to the AP via a helper. Following \cite{Liu2007}, we let node $h$ help node $k$ if
\begin{align}
\frac{1}{R_k}>\frac{1}{R_{kh}}+\frac{1}{R_h}\label{eq:helper2}
\end{align}
i.e., transmitting from $k$ via $h$ to the AP is better than transmitting directly from $k$ to the AP. The node $h$ is the best helper if
\begin{align}
h=\argmin_{l\in [1,N]}\frac{1}{R_{kl}}+\frac{1}{R_l}\label{eq:helper1}
\end{align}
We assume that node~$k$ knows the rate~$R_k$ and if it has a helper $h$ according
to \eqref{eq:helper2}, it also knows $R_{kh}$. We have a quasi-static environment in mind where a part of the nodes permanently experiences a
channel much worse than other nodes. We therefore assume that the rates of the
links remain constant over the period of interest. All nodes are restricted to the same fixed transmit power $\pwr$ during transmission.

\section{Throughput/Bit-Cost Tradeoff}
\label{sec:keyidea}
In this section, we motivate our investigations by a simple example that illustrates the importance of taking energy expenses into account when increasing throughput by cooperation.
\subsection{Throughput and Bit-Cost}
\label{subsec:throughputBitCost}
The \emph{throughput} $\thrp_k$ of node $k$ is the average amount of data bits per time that node $k$ successfully transmits. Only data belonging to $k$ is taken into account; data that $k$ forwards for other nodes does not contribute to the throughput $\thrp_k$. Let $\bar{\pwr}_k$ denote the \emph{average} power of node $k$ ($\bar{\pwr}$ is given by $\mathrm{transmit\,power\,} \pwr\times \mathrm{transmission\,time}/\mathrm{overall\,time}$). In contrast to the throughput $\thrp_k$, power spent while forwarding data of other nodes \emph{does} contribute to $\bar{\pwr}_k$. We define the \emph{bit-cost} $\btcs_k$ of $k$ as
\begin{align}
\btcs_k=\frac{\bar{\pwr}_k}{\thrp_k}
\end{align}
i.e., it measures the average amount of energy that node $k$ has to spend to successfully transmit one own data bit.

For exposition and comparison, we consider in this section \emph{Round Robin} as a centralized time division multiple access (TDMA) strategy. In a network of $N$ nodes scheduled with Round Robin, the nodes transmit one after each other in a circular order. Denote by $s_k$ the \emph{travel time} of one bit of node $k$ and denote by $t_k$ the \emph{transmission time} of node $k$, i.e., the overall time during which node $k$ is transmitting in one round. If node $k$ is transmitting directly to the AP and does not forward data of other nodes, then $s_k=t_k=1/R_k$. If node $k$ is transmitting directly to the AP and is forwarding data of the number of $H_k$ other nodes per round, then $s_k=1/R_k$ and $t_k=(H_k+1)/R_k$. If node $k$ transmits via node $h_k$, then $s_k=1/R_{kh_k}+1/R_{h_k}$ and $t_k=1/R_{kh_k}$. Throughput and bit-cost of node $k$ are thus given by
\begin{align}
\thrp_k = \frac{1 \mathrm{bit}}{\sum_{l=1}^N s_l}
\quad
\text{and}
\quad
\btcs_k = \frac{\bar{\pwr}_k}{\thrp_k}=\frac{\frac{t_k \pwr}{\sum_{l=1}^N s_l}}{\frac{1 \mathrm{bit}}{\sum_{l=1}^N s_l}}=\frac{t_k \pwr}{1 \mathrm{bit}}\label{eq:RRperformance}
\end{align}
Note that $\thrp_k=\thrp_l$ for all $k,l=1,\dotsc,N$. We can thus omit the index and simply refer to throughput $\thrp$, but we have to keep in mind that $\thrp$ is the throughput per node and not the throughput sum over all nodes in the network.

\subsection{A Toy Example}
\begin{figure}
\centering
\footnotesize
\psfrag{node1}{$n_1$}
\psfrag{node2}{$n_2$}
\psfrag{node3}{$n_3$}
\psfrag{node4}{AP}
\psfrag{1}{$1$}
\psfrag{3}{$3$}
\includegraphics[width=0.3\textwidth]{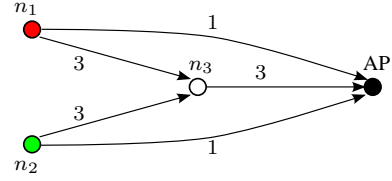}
\caption{A simple network with $3$ nodes and one AP. According to \eqref{eq:helper2}, node $n_3$ is a potential helper for both node $n_1$ and node $n_2$.}
\label{fig:toyExample}
\end{figure}
We now consider the simple network displayed in Figure~\ref{fig:toyExample}. Three nodes $n_1$, $n_2$, and $n_3$ want to transmit to the same AP. All nodes use the transmit power of $\pwr=1$~W. The rates are
\begin{align}
R_{n_1}=R_{n_2}=1\,\frac{\text{bit}}{\text{s}},\quad R_{n_1n_3}=R_{n_2n_3}=R_{n_3}=3\,\frac{\text{bit}}{\text{s}}.
\end{align}
\begin{figure*}
\centering
\subfloat[Flowchart for source node.]
{
\includegraphics[height=0.44\textwidth]{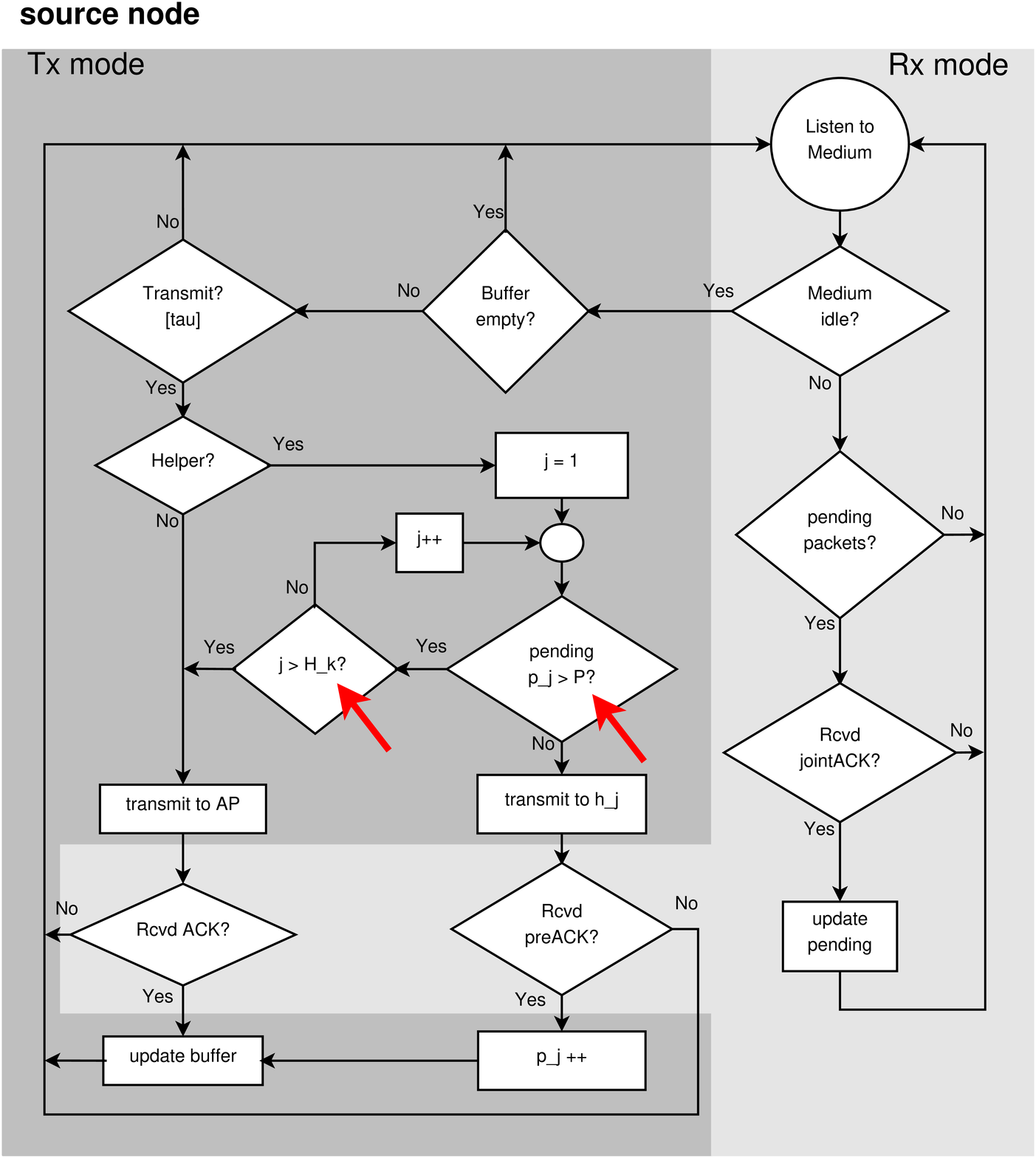}
\label{fig:source}
}
\hfill
\subfloat[Flowchart for helper node.]{
\includegraphics[height=0.44\textwidth]{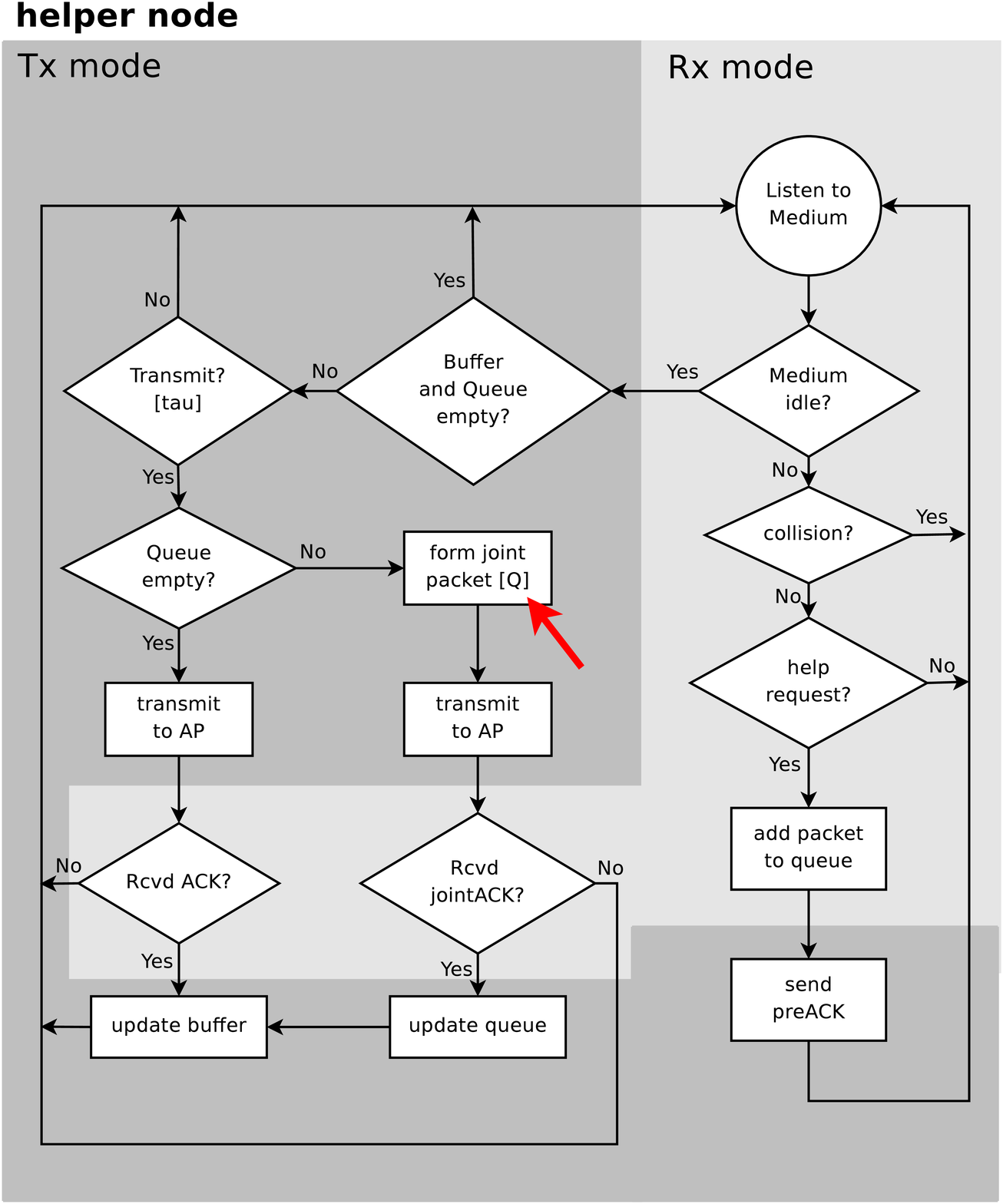}
\label{fig:helper}
}
\hfill
\subfloat[Flowchart for access point.]{
\includegraphics[height=0.44\textwidth]{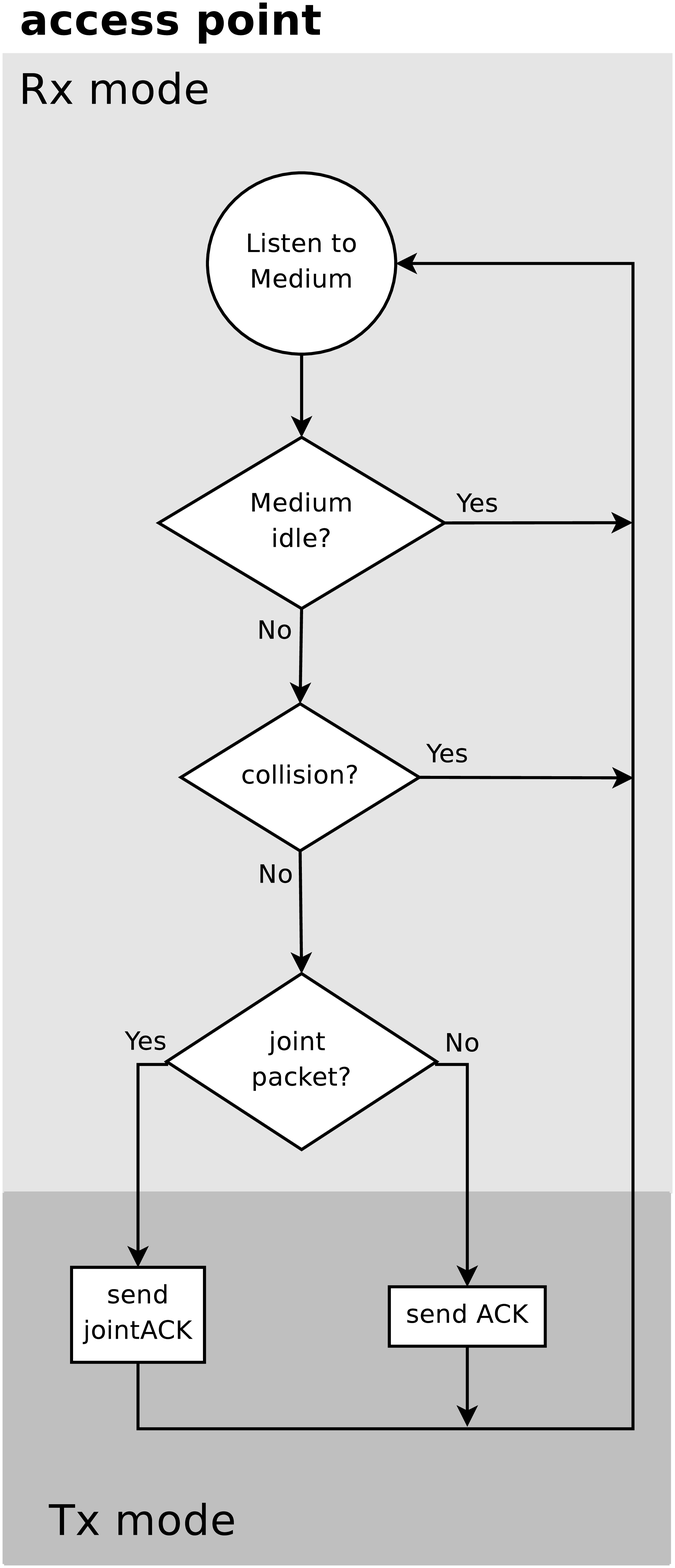}
\label{fig:ap}
}
\caption{Flowcharts for fairMAC as described in Subsection~\ref{subsec:fairmac}. The parameters maximum number $H_k$ of helpers, maximum number $P$ of pending packets per helper, and the maximum number $Q$ of packets forwarded at a time are emphasized by an arrow.}
\label{fig:flowcharts}
\end{figure*}
For simplicity, we omit units in the following. Because of $1/3+1/3<1$, according to \eqref{eq:helper2}, $n_3$ is a potential helper for both $n_1$ and $n_2$. For clear exposure, we postpone distributed scheduling through random access to the following sections \ref{sec:protocol} and \ref{sec:simulation} and schedule transmissions through Round Robin. The nodes $n_1$, $n_2$, and $n_3$ transmit one at a time in the fixed order $n_1,n_2,n_3,n_1,n_2,n_3,\dotsc$. In Direct Link, each node transmits one bit at a time directly to the AP, which takes the travel time $1$ for nodes $n_1$ and $n_2$ and the travel time $1/3$ for node $n_3$. In CoopMAC, nodes $n_1$ and $n_2$ first transmit their bits to $n_3$, which takes the time $1/3$. After receiving a bit from $n_1$ or $n_2$, node $n_3$ immediately forwards the received bit to the AP, which again takes the time $1/3$. Thus, the travel time in CoopMAC for bits of $n_1$ and $n_2$ is $1/3+1/3=2/3$ and for $n_3$, it is $1/3$. We can now use \eqref{eq:RRperformance} to calculate throughput and bit-cost of Round Robin based Direct Link and CoopMAC. For Direct Link, we get
\begin{align}
\thrp^\mathrm{dir} = \frac{1}{\frac{1}{1}+\frac{1}{1}+\frac{1}{3}}=\frac{3}{7},\quad\btcs^\mathrm{dir}_{n_1} = \btcs^\mathrm{dir}_{n_2} = 1,\quad \btcs^\mathrm{dir}_{n_3} = \frac{1}{3}.
\end{align}
For CoopMAC, we get
\begin{align}
\thrp^\mathrm{coop} = \frac{1}{\frac{2}{3}+\frac{2}{3}+\frac{1}{3}}=\frac{3}{5},\quad \btcs^\mathrm{coop}_{n_1} = \btcs^\mathrm{coop}_{n_2} = \frac{1}{3},\quad \btcs^\mathrm{coop}_{n_3} = 1.
\end{align}
As we can see, cooperation increases throughput from $3/7$ to $3/5$ and decreases the \emph{average} bit-cost from $7/9$ to $5/9$. However, the bit-cost of the helping node $n_3$ increases because of cooperation from $1/3$ to $1$: 
\begin{quote}
\emph{From the perspective of the helping node $n_3$, there is a tradeoff between throughput and bit-cost.}
\end{quote}

\section{fairMAC Protocol}
\label{sec:protocol}
For sake of clarity, we make some simplifying assumptions
for the MAC layer. Since we are
interested in equal throughput for all nodes, we assume that all nodes
operate in saturation mode, i.e., they are backlogged and we do not need to
consider packet arrival processes in our analysis. Under this assumption, it was
shown in \cite{Bianchi2000} that the DCF of IEEE 802.11 is equivalent to a
slotted carrier sense multiple access protocol (CSMA) with the two parameters slot length $\sigma$ and transmit probability $\tau$. We therefore base our protocols directly on slotted CSMA, which significantly simplifies presentation and comparison. In wireless networks, there are several reasons for packet losses. We include in
our work packet losses because of interference (collision) but neglect
other forms of packet losses. We further assume that control headers and acknowledgments (ACK) are transmitted at a base rate and that they can be decoded by all nodes in the 
network. To remain general, we assume that data packets are large enough such
that the specific size of control data is negligible. Finally, we assume
that ACKs never get lost.
\subsection{Reference Protocols}
We start by defining the two reference protocols Direct Link and CoopMAC. The definitions are identical to those given in \cite{Liu2007} but are included here to make the following more comprehensible. 
\subsubsection{Direct Link \cite{Bianchi2000}}
\label{subsec:directlink}
When node $k$ seeks to transmit a packet, it competes for the medium according
to CSMA:~if $k$ senses the channel idle in time slot $m$, it initiates a
transmission with probability $\tau$ in time slot $m+1$. If no other node is
transmitting at the same time, the AP can decode the packet and sends an ACK in return.
Otherwise, a collision occurs; no ACK is sent by the AP; node $k$ declares its
packet lost and will try to transmit again the same packet later.
\subsubsection{CoopMAC in base mode \cite{Liu2007}}
\label{subsec:coopmac}
All nodes initiate the transmission of an own packet in the same way as in
Direct Link. Assume that node $k$ initiates a transmission. We have to
distinguish two situations.
\begin{itemize}
\item Node $k$ has no helper. The transmission is performed according to the
Direct Link protocol.
\item Node $k$ has a helper $h$. In this case, $k$ transmits its packet to $h$
at rate $R_{kh}$. If $h$ can decode the packet, it immediately forwards the
packet to the AP at rate $R_{h}$. The AP sends an ACK to $k$. If $h$ cannot
decode the packet because of collision, it remains idle. Node $k$ detects the collision by not receiving the ACK. Node $k$ declares its packet lost and
tries to transmit the same packet again via $h$ later.
\end{itemize}
\subsection{fairMAC}
\label{subsec:fairmac}
CoopMAC was designed to maximize throughput. However, the resulting
energy expenses of potential helping nodes can become very large. Although a node addressed for help can in principal refuse to help, energy control at helping nodes is not incorporated in CoopMAC. This is because source $k$ decides when helper $h$ has to help: $h$ forwards immediately the packet from $k$. In fairMAC, this decision is taken by $h$: node $h$ stores the data from $k$ and
transmits it in conjunction with one of his own future packets.
The degree of cooperation in the network is in fairMAC controlled through the new parameter $Q$ at the helping nodes and the two new parameters $P$ and $H$ at the source nodes. All parameters will be explained in the following. The parameter $H$ determines from how many helping nodes a source node will demand help. For sake of clarity, we start by describing fairMAC when $H=1$, i.e., when each source node uses only the best helper according to \eqref{eq:helper2} and \eqref{eq:helper1}:
\subsubsection{fairMAC, one helper}
Helper node $h$ manages an additional, infinite packet queue for the packets to be forwarded. When $h$ receives a packet from $k$, $h$ adds it to this queue and notifies $k$ by sending a ``preACK'' to $k$. When node $h$ initiates a transmission
to the AP, it forms a joint packet consisting of own data from its buffer and data of up
to $Q$ packets from the forwarding queue. If there is no collision, the AP
successfully decodes the joint packet and sends one ``jointACK'' to $h$ and all other nodes with data in the joint packet. Node $h$ receives
the jointACK and removes the corresponding packets from the forwarding queue.

Source node $k$ tracks the packet delay at helper $h$ by a state
variable $p$ that indicates the number of pending packets. Each time $k$
transmits a packet to $h$ and receives a preACK, it increases $p$ by one.
When $p$ passes the maximum number of pending packets $P$, $k$ directly transmits its current packet to the AP. When $k$ receives a jointACK from the AP, it decreases $p$ by the number of its pending packets that helper $h$ finally forwarded to the AP in the corresponding joint packet.
\subsubsection{fairMAC, more than one helper} Each source node $k$ maintains a list of $H_k\leq H$ potential helpers. The set of helpers $\{h_1,\dotsc,h_{H_k}\}$ is ordered according to the quality of help provided, i.e., 
\begin{align}
h_l=\argmin_{j\in[1,N]\setminus\{h_1,\dotsc,h_{l-1}\}}\frac{1}{R_{kj}}+\frac{1}{R_j}.
\end{align}
Note that the sets of helpers are in general different for different source nodes $k$, but we omit to indicate this explicitly by an additional superscript $k$ for notational convenience. For the $l$th helper, node $k$ tracks the number of unacknowledged packets by the state variable $p_l$. Node $k$ tracks the helping node currently in use by the state variable $j$. Initially, $j=1$ and $k$ tries to transmit its packets via the best helper $h_1$. If $p_1>P$, node $k$ increases $j$ by one and tries to transmit via the second best helper and so on, i.e., node $k$ always tries to transmit via the best helper $h_l$ with $p_l\leq P$. If $p_l>P$ for all $l=1,\dotsc,H_k$, node $k$ uses direct transmission to the AP until $p_l\leq P$ for some $l$ after the reception of a corresponding jointACK. For a visualization of fairMAC, we provide flow charts in Fig.~\ref{fig:flowcharts}.

Analytical formulas for throughput and bit-cost of Direct Link, CoopMAC, and fairMAC with $H=1$ are given in \cite{Bocherer2010}.

\section{Numerical Results and Discussion}
\label{sec:simulation}
\begin{figure}
\centering
\footnotesize
\includegraphics[width=\columnwidth]{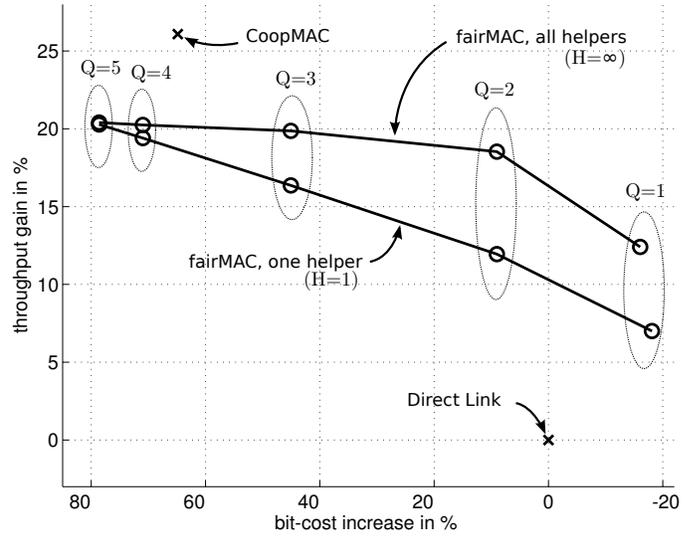}
\caption{Throughput gain and bit-cost increase compared to Direct Link for a network with $32$ nodes that are uniformly distributed in a unit circle. The normalized network parameter are $\tau=0.004$ and $\sigma=0.0088$. See \cite{Bocherer2010} for a detailed discussion of the applied normalization. All nodes use the same power $\pwr$ during transmission. We chose $\pwr$ such that the SNR of the node farthest away from the AP is $0$~dB at the AP. The maximum number of pending packets is constantly set to $P=10$. In the lower curve, fairMAC uses one helper per source node ($H=1$) and the number of packets forwarded at a time varies from $Q=1$ to $Q=5$. In the upper curve, $H$ is set to infinity and $Q$ varies over the same values. For comparison, the resulting performance of the full-cooperative protocol CoopMAC and of the reference strategy Direct Link are displayed. The results are obtained from a Monte Carlo simulation where the nodes compete 16 million times for the channel.}
\label{fig:thirtyNodes}
\end{figure}
We implement the protocols defined in Section~\ref{sec:protocol} in a custom network simulator in MATLAB. The implementation of fairMAC is based on the flowcharts from Fig.~\ref{fig:flowcharts}.

For a topology with $32$ nodes uniformly distributed in the unit circle, we calculate the transmission rates in the following way: we assume additive complex Gaussian noise of unit variance and complex Gaussian codebooks. The channel state is invariant over the period of interest and known both to sender and receiver. In this setting, we identify the achievable rate $R$ between sender and receiver with the mutual information between sent and received signal. It is given by
\begin{align}
R=\log(1+\mathrm{SNR})\quad\left[\text{nat}/\text{s}/\text{Hz}\right]
\end{align}
where $\log$ denotes the natural logarithm and where SNR denotes the signal-to-noise-ratio at the receiver. The SNR is equal to the transmission power $\pwr$ times an attenuation factor, which is given by the Euclidean distance between sender and receiver to the power of $\gamma$. We assign $\gamma=-3$ here. The value of $\pwr$ is chosen such that the SNR of the node in the network that is farthest away from the AP is equal to $0$~dB at the AP.
\subsection{Throughput/Bit-Cost Tradeoff}
We start by looking at how throughput changes when we vary the degree of cooperation. To this end, we choose the following parameters in fairMAC. We set the maximum number of helpers $H$ constantly equal to one and vary the number of packets $Q$ that each helper forwards at a time from $1$ to $5$. The number of pending packets $P$ is constantly set equal to $10$. We evaluate the achieved throughput versus the maximum bit-cost, where the maximum is taken over all nodes. The resulting curve is displayed in Figure~\ref{fig:thirtyNodes}. When the helping nodes only forward one packet at a time ($Q=1$), throughput is increased and bit-cost is decreased compared to Direct Link. However, when we increase $Q$ further, there is a tradeoff between throughput and bit-cost: the throughput increases further, however, the bit-cost also increases. In our example, for $Q=5$, the maximum is reached, i.e., increasing $Q$ further leaves throughput and bit-cost unchanged. As we can see in the figure, CoopMAC achieves a better throughput with lower bit-cost than fairMAC at $Q=5$. The reason is the following: while a helping node in CoopMAC immediately forwards a packet after reception, fairMAC accumulates for $Q=5$ up to five packets plus its own and then jointly transmits. This joint packet is very long. On the other hand, the source nodes only transmit their own (short) packets. As a result, long packets can collide with short packets. This is suboptimal and degrades the performance of CSMA.
\subsection{Throughput/NSI Tradeoff}
To evaluate the influence of NSI, we now set the number $H$ of helpers to infinity, i.e., each node knows all nodes that fulfill \eqref{eq:helper2}. For the other parameters, we use the same values as before, i.e., $P=10=\mathrm{const}$, $Q=1,2,3,4,5$. The resulting values are displayed in the upper curve in Figure~\ref{fig:thirtyNodes}. As we can see, more NSI increases throughput and decreases bit-cost over the whole range of the degree of cooperation $Q$: for each point on the $H=1$ curve, there is a point on the $H=\infty$ curve that has higher throughput and lower bit-cost. For increasing degree of cooperation $Q$, the $H=1$ curve converges to the $H=\infty$ curve and reaches it for $Q=5$. The reason is the following: although the source nodes know all potential helpers, the degree of cooperation is for $Q=5$ high enough to allow each source node to use the best helper in all transmissions.
\subsection{Application: Increasing Lifetime of a Network}
\begin{figure}
\centering
\footnotesize
\includegraphics[width=\columnwidth]{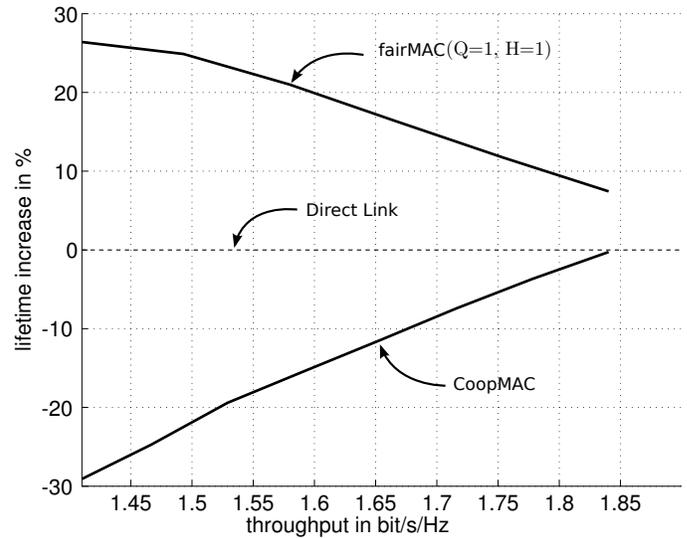}
\caption{The lifetime increase resulting from cooperation is plotted versus the achieved effective throughput. For the same throughput objective, full-cooperation (CoopMAC) decreases lifetime compared to Direct Link while partial cooperation increases lifetime. Partial cooperation is achieved by limiting the number of forwarded packets in fairMAC to $Q=1$. The NSI is set to $H=1$, i.e., each source node has only the knowledge of one helper. With this minimum amount of NSI that enables cooperation, a lifetime increase of over $25\%$ can be observed. Higher throughput is achieved by increasing the SNR. The benefits of two-hop decrease with higher SNR, which coincides with the theoretical results found in \cite{Bocherer2008}.}
\label{fig:lifetime}
\end{figure}

We now use the insights from the previous subsections to design a low complexity cooperative protocol that increases the lifetime of a wireless network. As reference strategies, we use Direct Link and CoopMAC. Since CoopMAC requires only the knowledge of the best helper, we choose $H=1$ for fairMAC, i.e., we set the amount of NSI to the minimum value that still allows cooperation. Since our objective is to maximize lifetime, we choose $Q=1$, i.e., the value that minimizes bit-cost according to Figure~\ref{fig:thirtyNodes}. We compare the strategies by plotting the lifetime against the observed effective throughput. For a fixed per-node energy budget $\mathsf{W}$, effective throughput $\mathsf{S}$, and maximum bit-cost $\mathsf{B}$, the network lifetime is given by $t=\mathsf{W}/(\mathsf{B}\cdot\mathsf{S})$. We vary the effective throughput by varying the SNR. As we can see in Figure~\ref{fig:lifetime}, the full-cooperative protocol CoopMAC \emph{decreases} lifetime over the whole range of effective throughput. On the other hand, fairMAC configured for partial cooperation \emph{increases} lifetime by up to $25\%$. For all considered protocols, the observed lifetime converges to the same value with increasing effective throughput because for high SNR, two-hop is not beneficial \cite{Bocherer2008}. CoopMAC actually reaches Direct Link: \eqref{eq:helper2} is not fulfilled anymore, and there are no potential helpers left in the network.

\bibliographystyle{IEEEtran}

\bibliography{IEEEabrv,confs-jrnls,Literatur}

\end{document}